\documentstyle[12pt,aaspp4,flushrt]{article}
\input{psfig}
\makeatletter
\@ifundefined{chapter}{\def\thebibliography#1{\section*{References\@mkboth
  {REFERENCES}{REFERENCES}}\list
  {\relax}{\setlength{\labelsep}{0em}
        \setlength{\itemindent}{-\bibhang}
        \setlength{\itemsep}{0pt}
        \setlength{\parsep}{0pt}
        \setlength{\leftmargin}{\bibhang}}
    \def\newblock{\hskip .11em plus .33em minus .07em}
    \sloppy\clubpenalty4000\widowpenalty4000
    \sfcode`\.=1000\relax}}%
{\def\thebibliography#1{\chapter*{Bibliography\@mkboth
  {BIBLIOGRAPHY}{BIBLIOGRAPHY}}\list
  {\relax}{\setlength{\labelsep}{0em}
        \setlength{\itemindent}{-\bibhang}
        \setlength{\itemsep}{0pt}
        \setlength{\parsep}{0pt}
        \setlength{\leftmargin}{\bibhang}}
    \def\newblock{\hskip .11em plus .33em minus .07em}
    \sloppy\clubpenalty4000\widowpenalty4000
    \sfcode`\.=1000\relax}}

\newlength{\bibhang}
\let\@internalcite\cite
\def\cite{\let\@citeleft(\let\@citeright)%
    \@ifstar{\citeyear}{\citefull}}
\def\citenp{\let\@citeleft\relax\let\@citeright\relax
    \@ifstar{\citeyear}{\citefull}}
\def\citefull{\def\astroncite##1##2{##1~##2}\@internalcite}
\def\citeyear{\def\astroncite##1##2{##2}\@internalcite}

\def\@citex[#1]#2{\if@filesw\immediate\write\@auxout{\string\citation{#2}}\fi
  \def\@citea{}\@cite{\@for\@citeb:=#2\do
    {\@citea\def\@citea{; }\@ifundefined
       {b@\@citeb}{{\bf ?}\@warning
       {Citation `\@citeb' on page \thepage \space undefined}}%
{\csname b@\@citeb\endcsname}}}{#1}}

\def\@cite#1#2{\@citeleft#1\if@tempswa , #2\fi\@citeright}
\def\@biblabel#1{}

\makeatother

\newcommand\approxlt{\mbox{$^{<}\hspace{-0.24cm}_{\sim}$}}

\begin{document}

\hfill{\small FERMILAB-Pub-99/004-A, CITA-99-1}

\title{Multiple-streaming and the Probability
Distribution of Density in Redshift Space} 
\author{Lam Hui\altaffilmark{1}, Lev Kofman\altaffilmark{2} and Sergei
F. Shandarin\altaffilmark{3}}
\altaffiltext{1}{NASA/Fermilab Astrophysics Center, Fermi
National Accelerator Laboratory, Batavia, IL 60510; e-mail: \it
lhui@fnal.gov}
\altaffiltext{2}{CITA, 60 St. George St., Toronto, M5S 3H8, Canada; e-mail: \it
kofman@cita.utoronto.ca}
\altaffiltext{3}{Department of Physics and Astronomy, University of
Kansas, Lawrence, KS 66045; e-mail: \it sergei@kusmos.phsx.ukans.edu}

\begin{abstract}
{We examine several aspects of redshift distortions by
expressing the redshift-space density
in terms of the eigenvalues and orientation of the local Lagrangian
deformation tensor. We explore the importance of multiple-streaming  
using the Zel'dovich approximation (ZA), 
and compute the average number of streams in both real and
redshift-space. It is found that multiple-streaming
can be significant in redshift-space but negligible
in real-space, even at moderate values of
the linear fluctuation amplitude ($\sigma_\ell \, \approxlt \, 1$).
Moreover, unlike their real-space counter-parts,
redshift-space multiple-streams can flow past each other
with minimal interactions. 
Such nonlinear redshift-space effects, which
are physically distinct from the fingers-of-God due to small-scale 
virialized motions, might in part explain the 
well-known departure of redshift distortions
from the classic linear prediction by Kaiser (1987), even at relatively
large scales where the corresponding density field in
real-space is well described by linear perturbation theory. 
We also compute using the ZA the probability distribution function (PDF) of
density, as well as $S_3$, in real and redshift-space, and
compare it with the PDF measured from N-body simulations.
The role of caustics in defining the character of the
high density tail is examined. It is found that (non-Lagrangian)
smoothing, due
to both finite resolution or discreteness and
small-scale velocity dispersions, 
is very effective in erasing caustic structures, unless
the initial power spectrum is sufficiently truncated.}
\end{abstract}


\section{Introduction}
\label{introduction}

The distortion of the density field in redshift-space by peculiar
motion is an old subject (see e.g.
\citenp{sargent77,bean83,davis83,kaiser87,hamilton92}).  
It is commonly held that on small scales, virialized motion causes the
stretching of 
structures along the line of sight, creating what is often called
the fingers-of-God, while on large scales, coherent
infall into high density regions causes compression, thereby making
structures appear more wall-like in the tangential directions than 
they are in real-space. The 
reader is referred to Hamilton \cite*{hamilton97} for a comprehensive review.

It has also been recognized for some time, however, that
the classic linear prediction by Kaiser
\cite*{kaiser87} overpredicts the compression, even at relatively
large scales where the corresponding density field in real-space
is quite linear ($\sigma_\ell \, \approxlt \, 1$). Conversely, 
the dilution effect embodied by the fingers-of-God appears to extend
to rather large scales. (See e.g.
\citenp{suto91,fisher93,gramann93,brainerd96,bromley97}.) 
We will refer to this phenomenon as translinear redshift distortions.

This effect has often been discussed in the context of attempts to measure
$\beta$ ($\Omega_m^{0.6}/b$, where $b$ is the bias) using the
ratio of the real-space to redshift-space two-point function or the
quadrupole-to-monopole ratio of the redshift-space power spectrum
(see e.g. \citenp{cole94}). It is found, both from observations,
as well as from N-body simulations, that the linear theory 
predictions for these ratios fail even at very large scales, where
perturbation theory is known to work well in real-space.
In general, one finds that the amount of line-of-sight squashing of 
the redshift-space 
two-point function is overpredicted by the linear perturbative
formulae.
The large uncertainties in published values of $\beta$ are at least in
part due to our poor understanding of this effect (see Table 1 of
\citenp{hamilton97}). 

A related phenomenon has also been observed in studies of one-point
statistics of the density field in redshift-space, albeit in
a slightly more subtle form. 
Hivon et al. \cite*{hbcj95} computed the
leading perturbative contribution to $S_3$ and $S_4$ in redshift-space. 
They observed that the agreement on large scales between the predicted values
and the measured values from N-body simulations is not as good
as it is in the case of real-space. Only at very small $\sigma_\ell$
(the linear 
fluctuation amplitude in real-space; in other words, very large
smoothing scales) did they obtain rough agreements. 
In fact, above some moderate $\sigma_\ell$ ($\sim 0.2$), the
value of $S_N$ in redshift-space generally lies below that in
real-space, suggesting an effect that resembles the fingers-of-God.

Recently, Taylor \& Hamilton \cite*{taylor96} and Fisher \& Nusser
\cite*{fn96} (see also Hatton \& Cole \citenp*{hc97} for detailed 
comparisons with simulations) revisited this issue by computing
the quadrupole-to-monopole ratio of the redshift-space power spectrum
using the Zel'dovich approximation (\citenp{za70}; henceforth as ZA). 
Both pairs of authors were able to obtain accurate predictions
for the shape of the quadrupole-to-monopole ratio on
translinear scales. In particular, they 
demonstrated analytically that the departure from the linear
prediction could take place at relatively large scales, and
they interpreted the success of the ZA in accounting for
this effect as an indication that the failure of the linear prediction
on translinear scales have more to do with coherent infall rather than
virialized motion. 

It is our aim here to continue this line of investigation of 
nonlinear effects in redshift distortions,
but we will focus our attention on one-point statistics.
Following earlier work by McGill \cite*{mcgill90}, we explore the
role of caustics and multiple-streaming in redshift-space.
It was pointed out by McGill \cite*{mcgill90}, who used
one-dimensional Zel'dovich dynamics, that redshift-space
caustics could form even when the real-space density field
is only mildly nonlinear. We will quantify this by computing
the average number of streams in both real and redshift-space
using the ZA. As will be shown, at $\sigma_\ell \sim 1$, the
degree of multiple-streaming could be significant in redshift-space
but negligible in real-space. This is covered in \S \ref{nostreams}.

The above finding indicates that the real to redshift-space mapping
enhances the level of nonlinearity. 
An interesting feature of multiple-streaming in redshift-space
is that, unlike their real-space counterparts, redshift-space
multiple streams to a first approximation can flow freely
past each other (at least before real-space caustics actually form), 
because they arise from physically distinct 
regions in real-space. This might provide an interesting explanation
for why the famous Kaiser formulae overpredicts the amount
of redshift-space compression along the line of sight,
even at moderate $\sigma_\ell$'s. Redshift-space multiple-streams
could then provide a mechanism for counteracting the linear compression.
The free-crossing of redshift-streams in the intermediate
regime is not unlike the familiar phenomenon of the thickening of
the ZA pancake. In this sense, the ZA, which can be viewed as a local
approximation where interaction is ignored (see \citenp{sz89} for a review;
see also \citenp{kp95,hb96}), is
well-suited for studying this phenomenon. 

The formalism for
computing the average number of streams was in fact
originally developed to
calculate the full one-point probability distribution function
(PDF hereafter) of density. Our work here is a natural extension of earlier work
by Kofman et al. \cite*{kofman94} who computed the PDF in real-space
using the ZA. We will also use the same methodology to compute 
the PDF as well as the skewness
$S_3$ in both real and redshift-space in the small $\sigma_\ell$
limit. The formalism is laid out in \S \ref{formalism}, the
computational method is explained in \S \ref{method}, and
the results of the PDF and $S_3$ calculation are discussed in
\S \ref{zpdf} and \ref{S3}.

We particularly focus our attention on the high density regime of the
PDF. Catastrophe theory tells us that caustics generally induce 
a $\rho^{-3}$ tail, which, however, is not observed in N-body
simulations with CDM-type (Cold-Dark-Matter) power spectra, either in
real or redshift-space.  
In fact, an intriguing property of the PDF in CDM
models is that it is well-fit by a single parameter lognormal
distribution (in both real and redshift-space), where the caustic-tail
is absent (for real space, see e.g. \citenp{kofman94,bk95}).
We explore the reason for it in \S \ref{smoothing}, and conclude that 
smoothing in redshift-space (or in real-space), due to both
finite resolution or discreteness and small-scale velocity
dispersions, plays an important
role in erasing the caustic-feature in the PDF  \footnote{
Smearing of caustics in the HDM model due to the thermal velocities
was considered by Zel'dovich \& Shandarin \cite*{ZS} and
Kotok \& Shandarin \cite*{KS}. }. 
To understand properly the behavior of the PDF then requires
a non-local calculation, which is beyond the scope of this paper.
We briefly discuss possible lines of attack in \S \ref{conclude}.

\section{Formalism}
\label{formalism}

For the purpose of this article, we denote by $\rho$ and $\rho_z$ 
the densities in real (or Eulerian) and redshift space respectively, normalized
by the mean density (i.e. $\langle\rho\rangle = \langle\rho_z\rangle =
1$). The evolution of $\rho$ is determined by the mapping from 
Lagrangian to Eulerian space, according to mass conservation:

\begin{equation}
\rho = \left| {\rm det}^{-1} \left[ {\partial x^i \over \partial q^j}
\right] \right|
\, ,
\label{rho}
\end{equation}
where ${x^i}$ and ${q^j}$ are the Eulerian and Lagrangian
coordinates. They are related by the displacement ${D^j}$
\begin{equation}
{\bf x} = {\bf q} + {\bf D} \, .
\label{displace}
\end{equation}

Similarly, the evolution of the redshift-space density is determined
by the mapping from Eulerian to redshift-space:
\begin{equation}
{\bf s} = {\bf x} + {{1+z}\over H} ({\bf v} \cdot {\bf e}) {\bf e} \, ,
\label{s}
\end{equation}
where ${\bf s}$ is the redshift-space coordinate, $z$ is the mean
redshift of interest, $H$ is the Hubble parameter
at redshift $z$, ${\bf v}$ is the peculiar velocity and ${\bf e}$ is
the unit vector along the line of sight.

The redshift-space density
$\rho_z$ is then given by \footnote{This is the single-stream
redshift-space density. See discussion at the end of the section.}
\begin{equation}
\rho_z = \left| {\rm det}^{-1} \left[ {\partial s^i \over \partial q^j}
\right]\right| = \rho \, \left| {\rm det}^{-1} \left[ {\partial s^i \over
\partial x^j} 
\right] \right| = \rho \left| 1 + {{1+z}\over H} {\partial {\bf
v}\over \partial 
x^i}\cdot{\bf e} e^i \right|^{-1} ,
\label{rhoz}
\end{equation}
where eq. (\ref{rho}) has been used for the second equality. 
We have used the remote-observer
approximation for the third equality, in other words, setting the term
$\partial {\bf e} /  
\partial x^i = 0 $. This is equivalent to assuming the
directional unit vector does not change appreciably for a small displacement
in the transverse direction. In addition, 
an interesting matrix identity has also been used:
${\rm det}[\delta_{ij} + A_i B_j] = 1 + {\bf A} \cdot {\bf B}$, where
${\bf A}$ and ${\bf B}$ are arbitrary vectors.

Let us now introduce the Zel'dovich approximation (ZA) to 
model the dynamics.

The ZA consists of assuming ${\bf D} \propto D_{+}$ where $D_+$ is the
linear time-dependent growth factor. Growing mode (in other words,
irrotational) initial condition then implies
\begin{equation}
{\bf D} = - D_+ {\bf \nabla_q} \psi \, ,
\label{psi}
\end{equation}
where $\psi$ is the displacement potential and ${\bf \nabla_q}$ in
component form is $\partial/\partial q^i$. 

The peculiar velocity is then simply ${\bf v} = {\bf \dot D}$, where
the derivative is with respect to conformal time
for a fixed ${\bf q}$. 

Putting the above into eq. (\ref{rhoz}), we arrive at
the following expression for $\rho_z$
\begin{equation}
\rho_z = \left| {\rm det}^{-1} {\bf X} \, \, \, (1 + f X^{-1}_{ik}
{D^k}_j e^i e^j)^{-1} \right| \, ,
\label{rhoz2}
\end{equation}
where
\begin{equation}
X_{ij} = \delta_{ij} + D_{ij} \quad , \quad D_{ij} = - D_{+}
{\partial^2 \psi \over {\partial q^i \partial q^j}} \, ,
\label{Xij}
\end{equation}
and $f \equiv \dot D a / (D \dot a) \sim \Omega_m^{0.6}$
\cite{peebles80},\footnote{A more accurate expression for $f$
can be found in \citenp{lahav91} and \citenp{carroll92}, but we 
will adopt $f \sim \Omega_m^{0.6}$, which is a very good approximation
at $z=0$.} with $a$ being the
scale factor and $\Omega_m$
being the fraction of critical density in matter.

Therefore, for a given $\Omega_m$, $\rho_z$ is uniquely determined 
by the line-of-sight unit vector $e^i$ and the displacement matrix
$D_{ij}$. The probability distribution of $\rho_z$ is then
dictated by the probability distributions of the later
two quantities.

For random orientations of the line-of-sight, the distribution of $e^i$
is given by 
\begin{equation}
P(e^i) d e^i = {\rm sin} \theta \, d\theta d\phi \, ,
\label{Pei}
\end{equation}
where $[e^1,e^2,e^3] = [{\rm sin}\theta {\rm cos}\phi, {\rm sin}\theta {\rm
sin}\phi, {\rm cos}\theta]$. 

For Gaussian random initial conditions, the probability distribution,
in Lagrangian space, 
of the displacement matrix $D_{ij}$ can be
computed exactly. One can always go to a frame in which $D_{ij}$
is diagonal, keeping in mind that the line-of-sight 
is randomly oriented with respect to the eigenvectors 
of this matrix. All we need then is the probability distribution
of its eigenvalues. 
We will denote this probability distribution in Lagrangian space
by $P (\lambda_j) d \lambda_{j}$, where $\lambda_j$'s ($j = 1, 2, 3$)
are the eigenvalues of $-D_{ij}$, $\lambda_1 \ge \lambda_2 \ge \lambda_3$.

Accordingly, eq. (\ref{rhoz2}) can be rewritten as 
\begin{eqnarray}
\rho_z =  | (1-\lambda_1) (1-\lambda_2) (1-\lambda_3) (1-
f  \sum_i {\lambda_i\over (1-\lambda_i)} (e^i)^2) |^{-1}
\label{rhoz3}
\end{eqnarray}
or further as
\begin{eqnarray}
\rho_z =&& |\, (1+f)(1-\lambda_1)(1-\lambda_2)(1-\lambda_3) \\ \nonumber
&& - f[(1-\lambda_2)(1-\lambda_3)e_1^2 + (1-\lambda_1)(1-\lambda_3)e_2^2
+ (1-\lambda_1)(1-\lambda_2)e_3^2] \, |^{-1}
\label{rhoz4}
\end{eqnarray}

%\begin{eqnarray}
%\rho_z = && | (1-\lambda_1) (1-\lambda_2) (1-\lambda_3) (1-
%f  \sum_i {\lambda_i\over (1-\lambda_i)} (e^i)^2) |^{-1} \\ \nonumber
%= && |\, (1+f)(1-\lambda_1)(1-\lambda_2)(1-\lambda_3) \\ \nonumber
%&& - f[(1-\lambda_2)(1-\lambda_3)e_1^2 + (1-\lambda_1)(1-\lambda_3)e_2^2
%+ (1-\lambda_1)(1-\lambda_2)e_3^2] \, |^{-1}
%\label{rhoz3}
%\end{eqnarray}

Thus, the density in redshift-space can ultimately be expressed
in terms of the eigenvalues and orientation of the local Lagrangian 
deformation tensor.
The statistical properties of $\rho_z$ can then be inferred from the 
well-known statistics of the deformation tensor.
Eqs. (\ref{rhoz3}), (\ref{rhoz4}) represents a redshift-space generalization of the 
familiar Zel'dovich 
formulae for the density in real-space. 
Just like the Zel'dovich result in real space, it sheds light
on the main features of the local structures in redshift space.
The first three factors in Eq. (\ref{rhoz3}) obviously arise from the
mapping from the Lagrangian to Eulerian (or real) space and the last factor comes
from that from real to redshift-space. The latter mapping 
strongly depends on the orientation of the deformation tensor
with respect to the line of sight ${\bf e}$. 

\begin{figure}[htb]
\centerline{\psfig{figure=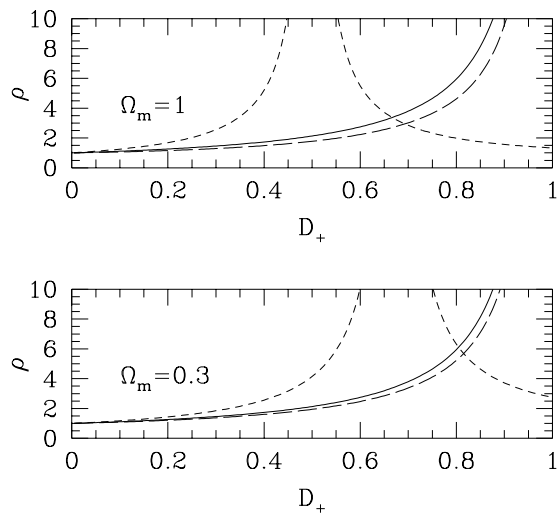,height=3.0in}}
\caption{The density evolution of a fluid element, as a function
of the growth factor $D_+$.
The eigenvalues of the deformation tensor (at $D_+ = 1$) are 
$\lambda_1 = 1, \lambda_2 = 0.5$ and   
$\lambda_3 = -0.5$. The solid line shows the evolution
of the real-space density. The short and long dashed lines
show how the redshift-space density evolves if the line
of sight aligns with $\lambda_1$ and $\lambda_3$ respectively.}
\label{sergei}
\end{figure}

In Fig. \ref{sergei}, we show an example of the density evolution of
of a fluid element, with the eigenvalues today ($D_+ = 1$) 
of $\lambda_1 = 1, \lambda_2 = 0.5$ and  
$\lambda_3 = -0.5$. Two cases are considered: when the line of sight is aligned
with the axis of fastest collapse ($\lambda_1$) (short dashed line), and
when the line of sight is aligned with the eigenvector associated with 
$\lambda_3$ (long dashed line). The evolution of the real-space density
is shown as a solid line. The real-space density formally diverges
at $D_+ = 1$ because of caustic-formation. The redshift-space density
tends to become nonlinear even earlier. 
In particular, this redshift-space amplification works most efficiently 
if the axis of the fastest collapse is aligned with the line of sight.
What we have here can be interpreted as a nonlinear manifestation
of the linear effect noticed by Kaiser (1987). 
Note how the redshift-space density decreases after (redshift-) caustic 
formation (at $D_+ = 0.5$, for the short dashed line).
This corresponds to the free-crossing of redshift-space multiple-streams,
which counteracts the Kaiser compression of structures along the line of sight.

It is also important to emphasize that eq. (\ref{rhoz3}) 
only gives us a local picture of the density distribution.
As in the familiar case of the ZA in real-space,  
our local Lagrangian treatment suggests an anisotropic structure, 
except that the structure is preferentially transverse to the line of sight.
However, it should be cautioned that the above equations tell
us nothing about the global distribution of these local structures,
whether they are aligned in a filament or a sheet is actually
determined by the correlation function of shear (see \citenp{bkp96}).
Moreover, small-scale velocity dispersions are likely to
at least smear out the caustics, which we will come
back to later. The only situation in which we can
say something about the global distribution of structures using
our local Lagrangian description is for a truncated initial power spectrum, 
in other words with a displacement field that is smoothed on large scales
in Lagrangian space, similar
to what is done in the truncated Zel'dovich approximation (\citenp{TZA1,TZA2}).

Let us now define two
different probability distribution functions (PDF) of density,
following the treatment of Kofman et al. \cite*{kofman94}.

The Lagrangian PDF of (redshift-space) density is:
\begin{equation}
P_{\rm L} (\rho_z) = \int \delta_D (\rho_z-{\rho_z}') P(\lambda_j) d\lambda_{j}
P(e^i) d e^i \, ,
\label{PL}
\end{equation}
where $\delta_D$ is the Dirac delta function, ${\rho_z}'$ in its argument
is a function of $\lambda_{j}$ and $e^i$ as expressed in eq.
(\ref{rhoz3}), $P(e^i) d e^i$ is given by eq. (\ref{Pei}), and 
$P(\lambda_j) d\lambda_{j}$ is the probability that a given Lagrangian
element has the set of eigenvalues $\lambda_j$'s falling in the
prescribed ranges. In other words, $P_{\rm L} (\rho_z) d\rho_z$ tells us the
fraction of the Lagrangian volume that has the redshift-space density
$\rho_z \pm d\rho_z/2$. 

To obtain the redshift-space PDF of the (redshift-space) density, two
effects need to be taken into account. 
The redshift-space PDF is supposed to tell us the fraction of the
redshift space volume that has the redshift-space density within
a particular range, say $\rho_z \pm d\rho_z/2$.
The first effect arises from the fact that 
the Lagrangian and redshift-space PDFs differ
by a multiplicative volume factor: a given Lagrangian volume
corresponds to a different redshift-space volume. This
factor is none other than $\rho_z$ itself. In other words,
the redshift-space PDF should be equal to the Lagrangian PDF
divided by $\rho_z$. 

\begin{figure}[htb]
\centerline{\psfig{figure=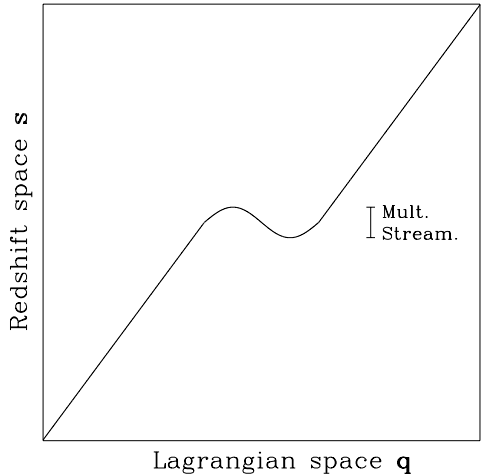,height=3.0in}}
\caption{A schematic diagram illustrating the mapping from
Lagrangian-space to redshift-space. Multiple-streaming occurs
in places where the mapping is not one-to-one, in other words,
several ${\bf q}$'s are mapped to the same ${\bf s}$.}
\label{qsmap}
\end{figure}

If there were no multiple-streaming, this would be the whole
story. The second effect that has to be taken into
account is illustrated schematically in Fig. \ref{qsmap}.
Multiple-streaming is said to occur in places where
the mapping from ${\bf q}$ to ${\bf s}$ is no longer
one-to-one. 

% qsmap.plot
%\begin{figure}[htb]
%\centerline{\psfig{figure=qsmap.ps,height=3.0in}}
%\caption{A schematic diagram illustrating the mapping from
%Lagrangian-space to redshift-space. Multiple-streaming occurs
%in places where the mapping is not one-to-one, in other words,
%several ${\bf q}$'s are mapped to the same ${\bf s}$.}
%\label{qsmap}
%\end{figure}

The expression for $\rho_z$ in eq. (\ref{rhoz}) is defined for
every Lagrangian coordinate ${\bf q}$. In other words, it is a single-stream
redshift-space density. In regions where multiple-streaming
occurs, the actual redshift-space density, at a particular ${\bf s}$
say, should be a sum of $\rho_z$'s over all ${\bf q}$'s that map onto
the same ${\bf s}$. Such a non-local calculation is beyond the scope of
this paper. Instead, following Kofman et al. \cite*{kofman94}, we
assume the fraction of the total redshift-space volume
that is occupied by these multiple-stream-regions is not large (i.e.
mildly nonlinear regime). However, there are inevitably some
such regions. A renormalizing factor $N_s$ is then needed when
computing the redshift-space PDF:
\begin{equation}
P_{\rm z} (\rho_z) = {N_s}^{-1} {\rho_z}^{-1} P_{\rm L} (\rho_z) \, ,
\label{Pz}
\end{equation}
where $P_{\rm L}$ is given in eq. (\ref{PL}) and 
\begin{equation}
N_s = \int_0^\infty {\rho_z}^{-1} P_{\rm L} (\rho_z) d\rho_z \, .
\label{Ns}
\end{equation}

The factor of ${\rho_z}^{-1}$ takes care of the first effect
mentioned above: it is the ratio of redshift-space volume to
Lagrangian-space volume. However, since ${\rho_z}$ is the
single-stream redshift-space density, an overcounting
of the total redshift-space volume over the total
Lagrangian-space volume occurs. The true ratio of the two
should be unity. $N_s$ gives
the ratio of the total {\it single-stream} redshift-space volume
to the total Lagrangian-space volume (the latter is also equal to
the true total
redshift-space volume with no overcounting), which is in general
larger than $1$ because of multiple-streaming (or, in other words,
overcounting). This quantity provides
the correct renormalizing factor for $P_{\rm z}$.
 
An additional bonus of the above formalism is that
we have a quantitative measure of the degree of multiple-streaming
in $N_s$. It tells us the average number of streams at
an arbitrary point in redshift-space. Fig. \ref{qsmap} provides
a nice illustration of how $N_s$ should be interpreted. Suppose $V^T$ is
the total true (no overcounting) volume 
in redshift-space (i.e. the ``volume'' shown on the y-axis). Suppose
further that $V^{1}$ is the part of $V^T$ that is in 
single-stream regions, and $V^{3}$ is the part that is in triple-stream
regions i.e. $V^T = V^{1} + V^{3}$. Recall that
$N_s$ is the ratio of the total {\it single-stream} redshift-space
volume to the total true redshift-space
volume, which means $N_s = (V^{1} + 3 V^{3}) / V^{T} = P(1) + 3
P(3)$, where $P(1)$ and $P(3)$ are the probabilities that a given region is
in single-stream and triple-stream regimes respectively. 
Hence, $N_s$ is exactly what one would refer to as the average
number of streams at a point in redshift-space. Obviously, the
argument extends to higher (but always odd) number of streams.

Note furthermore that while eq. (\ref{Pz}) gives
only an approximate redshift-space PDF in the limit of
little multiple-streaming ($N_s$ close to 1), eq. (\ref{Ns}) 
is an exact expression for the average number of streams, at least
within the framework of the ZA.
This should be kept in mind in the discussion of the next 
section.

Lastly, note that the above expressions (eq. [\ref{rhoz3}],
[\ref{PL}], [\ref{Pz}], [\ref{Ns}]) reduce to their real-space
counterparts in the limit $f = 0$, in other words, when
the real to redshift-space mapping is trivial. 

\section{The Average Number of Streams and the PDF: Method and Results}
\label{methodresults}

\subsection{Method}
\label{method}

The quantities we are interested in are $P_{\rm z} (\rho_z)$ and $N_s$
in eq. (\ref{Pz}) and (\ref{Ns}). Both of them require performing
the integral in eq. (\ref{PL}). 
Kofman et al. \cite*{kofman94} succeeded in reducing the analog of
this integral in real-space (i.e. setting $f = 0$ in eq.
[\ref{rhoz3}]) to a one-dimensional integral which has to be done
numerically. Redshift-space distortions introduce extra complications.
We will instead perform the integral in eq. (\ref{PL}) numerically 
right from the beginning. The delta function 
restriction $\delta_D[ \rho_z - {\rho_z}' (\lambda_j , e_i)]$ is most
easily dealt with using Monte Carlo techniques (see \citenp{rm95}
for a similar calculation in the context of the PDF of the optical depth
rather than the density in redshift-space).
In other words, we generate realizations of $\lambda_j$'s and
$e^i$'s, and identify combinations of them which satisfy 
the delta function restriction. 
The directional unit vector $e^i$ is easily handled
because it is randomly distributed over the 2-sphere, with the
probability distribution given by eq. (\ref{Pei}). 
On the other hand, the probability distribution for the $\lambda_j$'s
is somewhat more complicated.

The probability distribution of the eigenvalues of the displacement 
matrix $D_{ij}$, for a Gaussian random field, was first calculated
by Doroshkevich  \cite*{dorosh70}. We adopt a form
for this distribution which is better suited for the Monte Carlo computation
outlined above (following \citenp{bj94}; see
also \citenp{rm95}).  Let $-\lambda_j$'s be the eigenvalues of the
displacement matrix 
$D_{ij}$, and let $\Delta$, $\gamma$ and $\alpha$ be variables
defined through $\lambda_j$'s as follows 
\begin{equation}
[\lambda_1, \lambda_2, \lambda_3] = {\Delta\over 2} {\rm cos \gamma}
[1,1,1] + {\Delta\over \sqrt{5}}{\rm sin}\gamma [{\rm
cos}({{\alpha+2\pi}\over 3}), {\rm
cos}({{\alpha-2\pi}\over 3}), {\rm
cos}({\alpha \over 3})] \, ,
\label{angles}
\end{equation}
where the $\Delta$ ranges from $0$ to infinity, and both $\gamma$ and
$\alpha$ are between $0$ and $\pi$. The above ranges enforce the
ordering $\lambda_1 \leq \lambda_2 \leq \lambda_3$. It can be shown
that for any given combination of $\lambda_i$'s in this order, 
there corresponds a unique set of $\Delta$, $\gamma$ and $\alpha$. 

The probability distribution of $\lambda_j$'s is given by (\citenp{bj94}):
\begin{equation}
P({\lambda_{j}}) d \lambda_{j} = \left[{1\over 16} \left({3\over {2
\sigma_\ell}}\right)^6  \Delta^4 {\rm exp}\left(-{\Delta^2 \over
2}\left[{3\over {2\sigma_l}}\right]^2\right) d \Delta^2 \right] \left[{8\over
3\pi} {\rm sin}^4\gamma d\gamma\right] \left[{1\over 2} {\rm sin}\alpha
d\alpha\right] \, .
%see GPA41 and Kofman42
\label{Plambda}
\end{equation}
where $\sigma_\ell$ is the root-mean-squared (rms) amplitude of the
linear real-space density fluctuation.

A nice feature of the set of variables $\Delta$, $\gamma$ and
$\alpha$ is that their joint probability distribution factorizes.
$\Delta^2$ is distributed like a gamma-deviate.
$8\, {\rm sin}^4\gamma /(3\pi) d\gamma$ can be integrated exactly
to give $d\eta/2$ where $\eta = 1 - 2\gamma/\pi + 4 \, {\rm sin}
2\gamma/(3\pi) - \, {\rm sin}4\gamma/(6\pi)$, so that $\eta$ is
distributed uniformly between $-1$ and $1$. Lastly, ${\rm
cos}\alpha$ is also uniformly distributed between $-1$ and $1$. 
Standard numerical methods exist for generating variables distributed
in the above manner (see \citenp{press92}). 

Note that the only free parameters in the above calculation
are $\sigma_\ell$ and $\Omega_m$, or more precisely, $f$.
The former controls the distribution of $\lambda_j$'s through
eq. (\ref{Plambda}), while the latter determines the degree
of redshift-space distortion (eq. [\ref{rhoz2}] or
[\ref{rhoz3}]).

\subsection{The Average Number of (Lagrangian) Streams}
\label{nostreams}

% /d/ursa/lhui/Home/Codes/Pdf/PdfNoLog/PdfNoLogMulti/N.plot
%\begin{figure}[htb]
%\centerline{\psfig{figure=N.ps,height=3.0in}}
%\caption{The average number of Lagrangian streams in redshift space
%($N_s$) versus the rms amplitude of 
%the linear (real-space) density fluctuation ($\sigma_\ell$), for three
%different 
%values of $\Omega_m$. The $\Omega_m = 0$ curve gives also the 
%average number of Lagrangian streams in real-space.}
%\label{avestream}
%\end{figure}

Fig. \ref{avestream} shows the result of the calculation outlined
above for $N_s$, the average number of Lagrangian streams, defined in eq.
(\ref{Ns}). This is shown as a function of $\sigma_\ell$ (eq.
\ref{Plambda}), the 
rms amplitude of the linear real-space density fluctuation, for
three different values 
of $\Omega_m$. The $\Omega_m = 0$ case corresponds to the average number 
of streams in real-space.\footnote{Fig. 1 of Kofman et al.
\cite*{kofman94} contains an error for the real-space average number
of streams. For the perturbative result in real-space, see also
Munshi, Sahni \& Starobinsky \cite*{munshi94}.}

Several points should be noted. First, redshift-space distortion, or
peculiar motion, has the effect of increasing the amount of multiple-streaming
over that in real-space. Second, a larger
$\Omega_m$ means more redshift-space distortion,
hence a higher average number of streams.

At $\sigma_\ell = 1$, the average number of streams in
real-space is 
very close to 1.0, while the average
number of streams in redshift-space 
is 1.1 and 1.2 for $\Omega_m = 0.3$ and $\Omega_m = 1$
respectively. This means multiple-streaming can be insignificant in
real-space, but going on in earnest in redshift-space.\footnote{
$1.2$ might not seem too different from $1.0$. However, it should be kept in
mind this means, at any given point in redshift-space, 
one finds on the average $1.2$ streams. In other words, the total
single-stream redshift-volume is $20 \%$ higher than the total true
redshift-volume.}
In other words, even at relatively small $\sigma_\ell$, when the
real-space density field is quite linear, redshift-space
distortion introduces inherently 
non-perturbative effects through multiple-streaming. 

The above result perhaps explains the success of perturbation theory,
which is by nature a single-stream theory, 
in estimating various statistics of the density field in real-space, 
at $\sigma_\ell \, \approxlt \, 1$ (Many references should be cited here,
see \citenp{jb95} and \citenp{b96} for short reviews. See also
\citenp{sf96a} and \citenp{roman97} for loop corrections which can be
important at large enough $\sigma_\ell$.). The same does not appear to
be true
for statistics of the redshift-space density field, however. For
instance, it is well-known that the linear-theory 
redshift distortion factor deduced by Kaiser \cite*{kaiser87} fails to
describe the quadrupole-to-monopole ratio of the redshift-space
power spectrum even at very large scales (see Hatton \& Cole
\citenp*{hc97} and
references therein for a recent discussion). 
This failure is an indication of the importance of 
non-perturbative effects such as those
due to multiple-streaming in redshift-space.

The fact that multiple-streaming in real-space is less severe than
multiple-streaming in redshift-space has another interesting
implication.
Recall the Lagrangian to real-space to redshift-space mapping:
${\bf q} \rightarrow {\bf x} \rightarrow {\bf s}$. 
The multiple-(Lagrangian)-streaming in
redshift-space arises not so much from the ${\bf q} \rightarrow {\bf x}$
mapping (otherwise, the average number of Lagrangian streams
in real-space would not be so close to $1$ for, say, $\sigma_\ell = 1$),
but rather from the ${\bf x} \rightarrow 
{\bf s}$ mapping. 

This is interesting because the multiple-streams
at a given ${\bf s}$ then mostly come from distinct places in real
space i.e. they do not interact strongly, unlike the case of
multiple-(Lagrangian)-streams at the same ${\bf x}$. 
One well-known failure of the Zel'dovich Approximation is that it does
not account for the gravitational interaction between different
Lagrangian streams around a
real-space caustic i.e. the different streams (unphysically) freely go past 
each other at a Zel'dovich pancake. 
Around a redshift-space caustic, however, the multiple streams can
(approximately) free-stream past each other because they are not
located at the same real-space location.
This might provide a mechanism for partially canceling the line-of-sight
compression of structures pointed out by Kaiser (1987).

\subsection{The PDF of Redshift-space and Real-Space Density}
\label{zpdf}

We compute the PDF of redshift-space density according to eq.
(\ref{Pz}) for $\Omega_m=1$ and $\Omega_m=0$, and for 2 different
$\sigma_\ell$'s: the result is shown in Fig. \ref{ZApdf}. 
The $\Omega_m=0$ curves correspond also to the PDFs of the real-space
density. 
We have chosen $\sigma_\ell$'s small enough so that the average number of
streams is close to $1$ (see Fig. \ref{avestream}). This is to ensure
the validity of 
the approximation involved in eq. (\ref{Pz}). 

%Use /d/ursa/lhui/Home/Codes/Pdf/Sim/Dir1/Rs8and12linear/ZApdf.plot
%The ZA-pdf data file are obtained using the pdf code
%under /Dir1/Rs0/ZALNpdf/pdf.f 
%
%**Older version of a similar plot can be found in 
%  /d/ursa/lhui/Home/Codes/Pdf/PdfNoLog/pdf.plot
%\begin{figure}[htb]
%\centerline{\psfig{figure=ZApdf.ps,height=3.0in}}
%\caption{The ZA prediction for the PDF of redshift-space ({\it solid
%line}) and real-space ({\it 
%dashed line}) density for two different amplitudes of the linear
%(real-space) density fluctuation, $\sigma_\ell = 0.13$ and
%$\sigma_\ell = 0.49$. $\Omega_m=1$ for the redshift-space curves.}
%\label{ZApdf}
%\end{figure}

It can be seen that peculiar motion has the
effect of making the PDF look more nonlinear in redshift-space
than it is in real-space.
This should be expected based on the Kaiser \cite{kaiser87} effect
alone: large scale coherent infall into initial density 
enhancements (in real-space) shifts such regions further out into the
high-density tail of the PDF,  
while at the same time creating lower-density voids thereby
making the PDF peak at lower densities.

At higher $\sigma_\ell$'s, translinear distortions, or
redshift-space multiple-streaming, would start
to cancel the Kaiser compression, and begin to cause dilution
around the initial density enhancements.
At still higher $\sigma_\ell$'s, a
qualitatively different dilution effect, the 
finger-of-God, kicks in due to the formation of virialized clusters.
This happens after multiple-streaming becomes important in real-space.
Unfortunately, the PDF as expressed in
(\ref{rhoz3}) is incapable of taking multiple-streaming (in either
redshift or real-space) properly into
account,
as we have explained in \S \ref{formalism}. Nonetheless, we will
attempt to probe into the regime of multiple-streaming by considering
the effect of caustics 
on the high $\rho_z$ tail of the PDF in \S \ref{smoothing}.

How good is the ZA-predicted PDF as a description of the true PDF?
We show in Fig. \ref{simpdfV} a comparison of the ZA PDF versus the
PDF measured from an 
$\Omega_m=1$ CDM
simulation, for two different output times, smoothed at the same
comoving scale. The linear CDM power spectrum has $\Gamma = 0.25$, and
the box size is $100 \, {\rm \, h^{-1} Mpc}$, simulated on a $128^3$ grid.
The reader is referred to the Hydra consortium
({\tt http://coho.astro.uwo.ca/pub/data.html}; see also \citenp{couchman95})
for further details
of the simulation.
The one free parameter we have to fix for the ZA
prediction is the linear fluctuation amplitude $\sigma_\ell$. 
It can be seen that, as expected, $\sigma_\ell$ is always smaller than the 
fluctuation amplitude of the evolved-and-smoothed density field, which
we call $\sigma_{\rm nl}$. We also show lognormal distributions which
reproduce the 
evolved fluctuation amplitude of the simulation outputs. In other
words, 
\begin{equation}
P_{\rm logn.} (\rho_z) = {1\over \sqrt{2\pi} \sigma_{\rm eff} \rho_z}
e^{-{1\over 2 \sigma_{\rm eff}^2} ({\rm
ln}\rho_z + {1\over 2} \sigma_{\rm eff})^2} \, ,
\label{Plognorm}
\end{equation}
where $\sigma_{\rm eff}$ is chosen to equal $\sqrt{{\rm ln}(1+\sigma_{\rm
nl}^2)}$. It is impressive how well it fits the N-body
data out to large $\rho_z$'s. The same phenomenon has been observed in
real-space for a similar 
cosmological model (\citenp{kofman94}). The error-bars are estimated
by dividing the box into 4 
different subsamples, and computing the dispersion.
It should be emphasized that we have purposefully chosen
outputs, with suitable smoothing, that are only mildly nonlinear,
so that the average number of Lagrangian streams is very close to $1$,
making eq. (\ref{Pz}) a good approximation.

%O.K. for figure simpdfV.ps, see the result of 
%/d/ursa/lhui/Home/Codes/Pdf/Sim/Dir1/Rs8and12linear/simpdfV.plot
%where the simulation is with smoothing scale 8 cell-sizes
%for outputs at 049 and 023 (Couchman notation), with output
%information in aaa and Sim023/aaa respectively. Use code
%simpdfVlinear.f .
%The ZA is using sigma_linear = slightly smaller than sigma_nonlinear
%with s_l = 0.13 matching s_nl = 0.34 
%and s_l = 0.35 matching s_nl = 0.49. Used code Dir1/Rs0/ZALNpdf/pdf.f
%
%**For other simulation results, see
%  /d/ursa/lhui/Home/Codes/Pdf/Sim/guide
%  and Dir2/ and Dir3/ for open and lambda models
%  Also, if want logarithmic binning, look at /Rs8and12/simpdfV.f code
%  simpdf.f (without the V) gives the simulation pdf with no error-bar.
%\begin{figure}[htb]
%\centerline{\psfig{figure=simpdfV.ps,height=3.0in}}
%\caption{Comparison of the ZA-predicted PDFs ({\it solid lines}) with
%the measured PDFs ({\it points with error-bars}) from CDM,
%$\Omega_m=1$, 
%N-body simulations. The {\it dotted lines} represent a lognormal fit to the
%N-body data. The upper panel is from a simulation output at $a =
%0.23$, and the lower panel is from an output at $a =
%0.49$, each smoothed with a Gaussian window of $6.25 \, {\rm h}^{-1} {\rm
%Mpc}$. The actual rms amplitudes of the density fluctuation
%($\sigma_{\rm nl}$) are
%$0.34$ and $0.49$ 
%respectively, while the linear $\sigma_\ell$'s used for the ZA-curves
%are $0.13$ and $0.35$.} 
%\label{simpdfV}
%\end{figure}

The overall agreement between the ZA PDF and the N-body data is decent
at these low values of $\sigma_{\ell}$. However, the ZA PDF tends to
overpredict the probability of high $\rho_z$. It is in fact well-known that
the ZA PDF has a long tail of $P(\rho_z) \propto 1/\rho_z^3$, 
due to the presence of caustics. This is true in both real and
redshift-space.
It implies that, unfortunately, $\sigma_{\ell}$ has to be left as a
free parameter in the above comparison 
because the ZA PDF does not yield a well-defined $\sigma_{\rm nl}$. 
We will revisit this issue in \S \ref{smoothing}.

\subsection{$S_3$ in Redshift-space and Real-space}
\label{S3}

The methodology outlined in \S \ref{method} can be easily adopted to compute
other one-point statistics, such as $S_3$:
\begin{equation}
S_3 = {\langle (\rho_z - 1)^3 \rangle \over \langle (\rho_z -1)^2
\rangle^2}
\label{S3def}
\end{equation}
As in the case of
$\sigma_{\rm nl}$, because of
the caustic-induced tail of the ZA PDF, $S_3$ is, strictly speaking, 
undefined except 
in the limit of vanishing $\sigma_\ell$. The finite number of
realizations used in our Monte Carlo method
provides a natural regularization by truncating at some high
$\rho_z$'s, thereby 
giving a finite $S_3$. As one decreases the input parameter
$\sigma_\ell$ in the Monte Carlo integration,
the output $S_3$ converges eventually to the correct ZA predicted value for
infinitesimal $\sigma_\ell$. Our results are summarized in Table
\ref{S3tab} (see also e.g. \citenp{bk95} for the real-space ZA prediction).
We find that the rate of convergence for $S_4$, $S_5$, etc
becomes progressively slower, and so we only give the $S_3$ values
here. 

It can be seen that peculiar motion tends to slightly increase $S_3$ over
its real-space value,
as the reader might have guessed from the PDFs shown in Fig.
\ref{ZApdf}. This behavior is consistent with what is found
by Hivon et el. \cite*{hbcj95} in the case of the exact dynamics,
in the small $\sigma_\ell$ limit. However, as is well-known in the
case of real-space, in the perturbative limit, the ZA systematically
gives lower $S_3$ than 
the exact dynamics.

As emphasized by
Hivon et al. \cite*{hbcj95}, the perturbative
prediction breaks down even at relatively small $\sigma_\ell$'s (say
$0.2$), 
or large smoothing scales, and the actual N-body $S_3$ is generally
{\it smaller} in redshift-space than in real-space. They
attributed it to some kind of finger-of-God effect on large
scales, which we view as in part due to multiple-streaming
in redshift-space (that is distinct from the familiar small-scale
finger-of-God). Unfortunately, the ZA yields a meaningful skewness
only in the vanishing $\sigma_\ell$ limit, making it
difficult to investigate the impact of these effects on $S_3$.

%\begin{table}
%\begin{center}
%\begin{tabular}{|ccc|}\hline
%$\sigma_\ell$ & $\Omega_m $ & $S_3$ \\ \hline
%0.01 & 0 & 4.0 \\
%0.01 & 0.3 & 4.1 \\
%0.01 & 1.0 & 4.2 \\ \hline
%\end{tabular}
%\end{center}
%\caption{\label{S3tab} The ZA predicted $S_3$ in real-space (the
%$\Omega_m = 0$ case)  
%and redshift-space, for $\sigma_\ell = 0.01$}
%\end{table}

\section{Erasing Caustics ?}
\label{smoothing}

Why does the ZA PDF predict a long high-$\rho_z$ tail, and
why is it not observed in N-body simulations, at least for
CDM-type models? 
We believe the answer to the latter at least in part has to do 
with smoothing due to finite resolution or discreteness as well as small-scale 
velocity dispersions.
But let us first briefly review
where the high $\rho_z$ tail comes 
from. The reader is referred to Shandarin \& Zel'dovich \cite*{sz89}
for a review. Most of our arguments below apply equally well to 
real-space caustics.

A caustic is located at a point where $\rho_z$ or ${\rm det}^{-1}
\partial s_i /\partial q_j$ diverges. As Zel'dovich \cite*{za70}
originally argued,
this is generically due to the vanishing of only one of the
eigenvalues of the matrix $\partial s_i /\partial q_j$. Hence,
we can focus our attention on the direction that aligns with the
eigenvector associated with the vanishing eigenvalue. We will denote
the relevant one-dimensional coordinates simply by $s$ and $q$. The situation
is depicted in Fig. \ref{qsmap}. 

Taylor expanding $s$ as a function of $q$ around $q = q_c$
(corresponding to $s = s_c$ where the caustic is), one can see that to
the lowest order, $s$ is a quadratic function of $q$ i.e.
\begin{equation}
s - s_c = {1\over 2} \left.{\partial^2 s \over \partial q^2}\right|_{q
= q_c} (q - q_c)^2
\label{sqc}
\end{equation}

In other words, $\rho_z$ should fall off as $1/ (q-q_c)$ or
$1/\sqrt{s-s_c}$, around the caustic. Note that, strictly speaking,
$\rho_z$ should be given by the sum over the contribution from each
multiple-stream around the 
caustic (as in the case depicted in Fig. \ref{qsmap}), but the
singularity due to $1/ (q-q_c)$ or $1/\sqrt{s-s_c}$ dominates. Note also that
this singular behavior of $\rho_z$ occurs only on one
side of the 
caustic in $s$-space. 

To understand the caustic's influence on the PDF, the crucial point 
to remember 
is that one has equal probability of 
locating at any position in $s$-space with respect to the location of
the caustic. Therefore, $P(s) ds \propto P(\rho_z) d
\rho_z$, with $P(s)$ independent of $s$, from which the
$P(\rho_z) \propto 1/\rho_z^3$ behavior follows. The premise is that
asymptotically high $\rho_z$'s occur only around caustics. 

Why is the caustic-induced feature of the PDF then not observed?

We believe the answer lies in final (as opposed to initial or
Lagrangian) smoothing. There are two sources
of final smoothing: one due to finite resolution or the discrete nature
of the simulations (or observations), and the other due
to smearing by small-scale velocity dispersions.
Two conditions have to be met for the caustic-induced high $\rho_z$ tail to
survive these kinds of inevitable smoothing in $s$-space. 

First, it is important that the scale of the caustic in $s$-space,
$\Delta s_{\rm caustic}$, is larger than the smoothing scale,
$\Delta s_{\rm smooth}$, where the smoothing scale can
be taken to be whichever is dominant of the two kinds of smoothing
mentioned above $\Delta s_{\rm smooth} = {\rm max} (\Delta s_{\rm
res.}, \sigma_v (1+z) / H)$. The smoothing scale $\Delta s_{\rm
res.}$ is associated with the finite final resolution, and
$\sigma_v$ is the velocity dispersion along the line of sight.

The scale of the caustic in $s$-space is related to
its counterpart in $q$-space by eq. (\ref{sqc}):
\begin{equation}
\Delta s_{\rm caustic} = {1\over 2} \left|{\partial^2 s \over \partial
q^2}\right|_{q_c} \Delta q_{\rm caustic}^2 \, ,
\label{Deltasc}
\end{equation}
where $\Delta q_{\rm caustic}$ is the scale of the caustic in $q$-space.
A good definition for $\Delta q_{\rm caustic}$ is given by
the magnitude of $q - q_c$ such that the next higher order term ignored in eq.
(\ref{sqc}) becomes significant. In other words, 
$\Delta q_{\rm caustic} \sim [\partial^3 s
/ \partial q^3]^{-1} [\partial^2 s / \partial q^2]$. 
One can use even higher order terms for its definition, but
it would not affect our arguments below.

Let us then express the first condition for the survival of the
caustic-induced feature as
\begin{equation}
\Delta s_{\rm smooth} < {1\over 2} \left|{\partial^2 s \over \partial
q^2}\right|_{q_c} \Delta q_{\rm caustic}^2 \, .
\label{condition1}
\end{equation}

The second condition has to do with the density profile of a smoothed
caustic. Assume that the first condition above is met, the smoothed
profile of the caustic would be given by $\rho_z (s) = f(s) /
\sqrt{|\partial^2 s /\partial q^2|_{q_c} \Delta s_{\rm smooth}}$ where
$f(s)$ is some dimensionless profile which, at its peak, is
equal to $O(1)$. This is a very generic prediction quite independent of
the form or origin of the smoothing.
This means the smoothed caustic peaks at ${\rho_z}_{\rm max} \sim
1 / \sqrt{|\partial^2 s /\partial q^2|_{q_c} \Delta s_{\rm smooth}}$. 
Now, if one would like such a smoothed caustic to contribute 
to the high $\rho_z$ tail, it is important that ${\rho_z}_{\rm max}
\gg 1$. Hence, the second condition for the survival of the
caustic-induced feature in the PDF is:
\begin{equation}
\left|{\partial^2 s \over \partial q^2}\right|_{q_c} \ll {1 \over \Delta s_{\rm
smooth}}
\label{condition2}
\end{equation}

The two conditions in eq. (\ref{condition1}) and (\ref{condition2})
together imply the necessary requirement that
\begin{equation}
{\Delta s_{\rm smooth} \over \Delta q_{\rm caustic}} \ll 1 \, .
\label{condition3}
\end{equation}

Therefore, the only remaining question is what the typical $\Delta
q_{\rm caustic}$ should be. This can be estimated using the definition
suggested before: that $\Delta q_{\rm caustic}$ is given by
$ [\partial^3 s
/ \partial q^3]^{-1}$ $[\partial^2 s / \partial q^2]$ evaluated at the
caustic. One can calculate exactly what the average value of this
ratio should be, for Gaussian initial conditions, imposing the
constraint that the first derivative vanishes. All one needs is
the joint probability distribution of all the derivatives up to the
third one {\it at a single point}. However, there is no need to do any
calculation. The only relevant scale in this problem is the 
{\it initial} smoothing scale of the density field (or the
displacement field) in $q$-space, let us call it 
$\Delta q_{\rm smooth}$. Therefore $\Delta q_{\rm
caustic} \sim \Delta q_{\rm smooth}$. 

For CDM-type power spectra, the initial smoothing scale $\Delta q_{\rm
smooth}$ is of the order of a few grid-spacings
(unless the initial power spectrum is deliberately
smoothed or truncated), while the final smoothing scale $\Delta s_{\rm
smooth}$ is at least
also a few grid-spacings (and probably even larger due to 
small-scale velocity dispersions). This implies the condition in eq.
(\ref{condition3}) is never met, hence explaining why the theoretical
caustic-induced tail of the PDF is not observed, at least in CDM
simulations. In other words, to understand the high-$\rho_z$ behavior
of the PDF properly, both multiple-streaming and smoothing have
to be properly taken into account.
This also points to a weakness of the calculation
presented in \S \ref{formalism}. Because final smoothing in $s$-space
is not taken into account in the formalism, effects like what is
discussed above are missed. We hope to pursue improvement of the
calculation in this direction in the future.

\section{Discussion}
\label{conclude}

\begin{figure}[htb]
\centerline{\psfig{figure=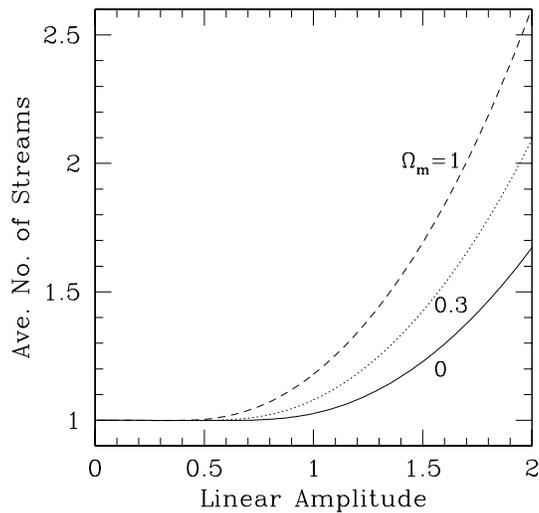,height=3.0in}}
\caption{The average number of Lagrangian streams in redshift space
($N_s$) versus the rms amplitude of 
the linear (real-space) density fluctuation ($\sigma_\ell$), for three
different 
values of $\Omega_m$. The $\Omega_m = 0$ curve gives also the 
average number of Lagrangian streams in real-space.}
\label{avestream}
\end{figure}

\begin{figure}[htb]
\centerline{\psfig{figure=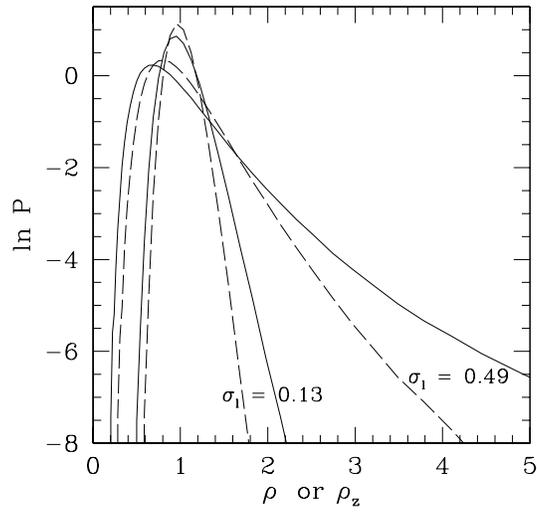,height=3.0in}}
\caption{The ZA prediction for the PDF of redshift-space ({\it solid
line}) and real-space ({\it 
dashed line}) density for two different amplitudes of the linear
(real-space) density fluctuation, $\sigma_\ell = 0.13$ and
$\sigma_\ell = 0.49$. $\Omega_m=1$ for the redshift-space curves.}
\label{ZApdf}
\end{figure}

Using the Zel'dovich approximation, we find that peculiar motion 
significantly raises the degree of multiple-streaming.
We quantify this by computing the average number of Lagrangian-streams
in both real and redshift-space. As is shown in Fig. \ref{avestream}, 
multiple-streaming can be insignificant in real-space, but
non-negligible in redshift-space, even at moderate values of
$\sigma_\ell$. This implies that most of the multiple-streaming
in redshift-space arises from the ${\bf x} \rightarrow {\bf s}$
mapping (real or Eulerian to redshift-space), rather than from the ${\bf
q} \rightarrow {\bf x}$ mapping (Lagrangian to Eulerian space).

That the level of nonlinearity is amplified by the real to redshift-space
mapping implies non-perturbative effects have to be taken into account
in the calculation of redshift distortions, even
on scales where the real-space density field is relatively linear.
An interesting feature of redshift-space multiple-streaming
is that redshift-space multiple-streams can freely flow past
each other, unlike their real-space counterparts.
This counteracts the familiar large scale compression of structures
along the redshift direction first
pointed out by Kaiser (1987), but it is physically distinct
from the stretching of structures due to virialized motions
on small scales.
This might offer an explanation of why, even on relatively large
scales, linear theory overpredicts the amount of reshift-space squashing
of the two-point function along the line of sight.
To make this quantitative, however, requires a detailed calculation,
which we will discuss briefly below.

\begin{figure}[htb]
\centerline{\psfig{figure=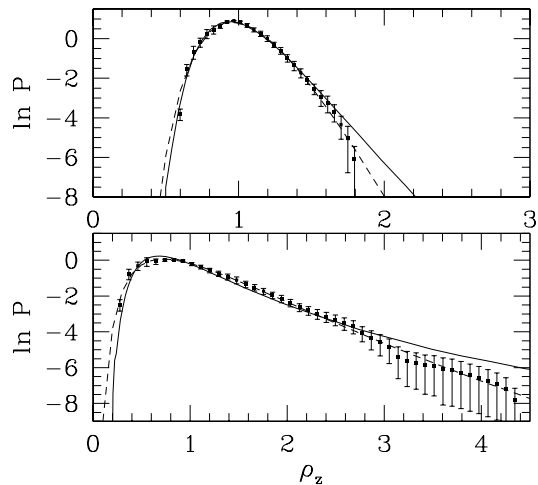,height=3.0in}}
\caption{Comparison of the ZA-predicted PDFs ({\it solid lines}) with
the measured PDFs ({\it points with error-bars}) from CDM,
$\Omega_m=1$, 
N-body simulations. The {\it dotted lines} represent a lognormal fit to the
N-body data. The upper panel is from a simulation output at $a =
0.23$, and the lower panel is from an output at $a =
0.49$, each smoothed with a Gaussian window of $6.25 \, {\rm h}^{-1} {\rm
Mpc}$. The actual rms amplitudes of the density fluctuation
($\sigma_{\rm nl}$) are
$0.34$ and $0.49$ 
respectively, while the linear $\sigma_\ell$'s used for the ZA-curves
are $0.13$ and $0.35$.} 
\label{simpdfV}
\end{figure}

We have computed also using the ZA the probability distribution function
(PDF) of density, 
as well as $S_3$,
in both real and redshift-space. 
At the largest scales, or small $\sigma_\ell$ limit,
$S_3$ does not appear to change significantly
from real to redshift-space, consistent with the finding of
Hivon et al. \cite*{hbcj95}.
We have compared the ZA PDF with that
measured from a CDM N-body simulation (Fig \ref{simpdfV}). We find that
the PDF in redshift-space is well-fit by a lognormal distribution,
similar to its counter-part in real-space (\citenp{kofman94,bk95}). 
This suggests the PDF
in both real and redshift-space can be described by a single-parameter
family of functions ($\sigma_{\rm eff}$ in eq. [\ref{Plognorm}]).
This is certainly worth exploring with more
N-body simulations and different cosmological models
(see \citenp{sg98}).

The simulation PDF reveals no caustic-induced 
high $\rho_z$ tail, contrary to the ZA prediction. 
The same has been observed before in real-space as well (\citenp{kofman94}). 
We argue in \S \ref{smoothing} that this is at least in part 
due to the fact that final
smoothing in redshift-space (or real-space) is not taken
into account in our formalism. Smoothing naturally arises
because of finite resolution or discreteness, as well
as due to small-scale velocity dispersions. We give the criterion for
the survival of the caustic feature through smoothing in eq.
(\ref{condition3}): the final-smoothing scale has to
be smaller than the Lagrangian-smoothing scale. 

In practice, we expect caustics to be efficiently erased
by final-smoothing, unless the initial power spectrum is
sufficiently smoothed or truncated.
This is reminiscent of a related situation in gravitational lensing:
that caustic features can only be observed if the source-size
is sufficiently small.

As mentioned before, a proper treatment of the statistics of
the density field
in redshift-space requires tackling the issues
of multiple-streaming and final-smoothing.
While such a calculation is beyond the scope of the present paper,
Taylor \& Hamilton \cite*{taylor96} and Fisher \& Nusser \cite*{fn96}
have already made significant progress in using the ZA to compute
the quadrupole-to-monopole ratio of the redshift-space power spectrum.
They use the following expression for the Fourier transform
of the redshift-space density:
\begin{equation}
\tilde \rho_z ({\bf k}) = \int d^3 q e^{-i {\bf k} \cdot ({\bf q} + {\bf D_s})}
\label{rhozk}
\end{equation}
where ${\bf D_s}$ as a function of ${\bf q}$ is the displacement
from the Lagrangian position ${\bf q}$ to the redsfhit-space position ${\bf
s}$. It comes from the following 
basic equation for the redshift-space density:
$\rho_z ({\bf s}) = \int d^3 q \delta_D ( {\bf s} - {\bf q} - {\bf D_s} )$,
which implicitly allows for multiple-streaming
by automatically summing over density contributions from all possible
streams.
The key here is that the mapping from real to redshift-space
is treated exactly, instead of perturbatively, hence
the non-perturbative nature of the redshift-space density
field is taken into accout even if the real-space
density field is still quite linear.
These authors find that the ZA predicts the right-shape for the
quadrupole to monopole ratio but underestimates the
zero-crossing scale (see also Hatton \& Cole \citenp*{hc97}).
One can take their treatment one step further by using
approximations that are closer to the exact dynamics (see e.g.
Scoccimarro \citenp*{roman98}). Whatever the approach, it
appears likely that, because of the prominence of multiple-streaming
in redshift-space, a pure perturbative calculation would be
inadequate. Non-perturbative effects such as those due to
multiple-streaming (or even pre-multiple-streaming) have to be taken into
account explicitly.

Lastly, our Lagrangian analysis here still leaves
open the question of how the local pancakes or caustics
are spatially distributed globally, whether they preferentially
lie along filaments or sheets. Calculations have shown that
filamentary structures in real-space are preferred for Gaussian initial
conditions (e.g. \citenp{bkp96}, \citenp{YS}). 
It remains to be seen how the redshift-space mapping might
alter this picture.
Our local calculation, however, does provide a valid description
of the global distribution of structures if the initial
power spectrum is sufficiently smoothed, as is done in
the truncated Zel'dovich approximation.

LH is grateful to Roman Scoccimarro for many useful discussions and
help with the simulations. The CDM simulations analyzed in this work
were obtained from the data bank of N-body simulations provided by the
Hydra consortium  
({\tt http://coho.astro.uwo.ca/ pub/data.html}) and produced using
the Hydra N-body code (\citenp{couchman95}). 
LH thanks Dick Bond and the Canadian Institute for Theoretical 
Astrophysics for hospitality and an excellent working environment.
This work was in part supported by the DOE and the NASA grant NAG 5-7092 at
Fermilab. SS acknowledges the support of EPSCoR 1998 grant, GRF grant at
UK, and thanks the Canadian Institute for Theoretical
Astrophysics for hospitality.

%\bibliographystyle{astro}
%\bibliography{pdff}

\clearpage

\begin{table}[htb]
\begin{center}
\begin{tabular}{|ccc|}\hline
$\sigma_\ell$ & $\Omega_m $ & $S_3$ \\ \hline
0.01 & 0 & 4.0 \\
0.01 & 0.3 & 4.1 \\
0.01 & 1.0 & 4.2 \\ \hline
\end{tabular}
\end{center}
\caption{\label{S3tab} The ZA predicted $S_3$ in real-space (the
$\Omega_m = 0$ case)  
and redshift-space, for $\sigma_\ell = 0.01$}
\end{table}

%lambda.plot

% qsmap3.plot

% /d/ursa/lhui/Home/Codes/Pdf/PdfNoLog/PdfNoLogMulti/N.plot

%Use /d/ursa/lhui/Home/Codes/Pdf/Sim/Dir1/Rs8and12linear/ZApdf.plot
%The ZA-pdf data file are obtained using the pdf code
%under /Dir1/Rs0/ZALNpdf/pdf.f 
%
%**Older version of a similar plot can be found in 
%  /d/ursa/lhui/Home/Codes/Pdf/PdfNoLog/pdf.plot

%O.K. for figure simpdfV.ps, see the result of 
%/d/ursa/lhui/Home/Codes/Pdf/Sim/Dir1/Rs8and12linear/simpdfV.plot
%where the simulation is with smoothing scale 8 cell-sizes
%for outputs at 049 and 023 (Couchman notation), with output
%information in aaa and Sim023/aaa respectively. Use code
%simpdfVlinear.f .
%The ZA is using sigma_linear = slightly smaller than sigma_nonlinear
%with s_l = 0.13 matching s_nl = 0.34 
%and s_l = 0.35 matching s_nl = 0.49. Used code Dir1/Rs0/ZALNpdf/pdf.f
%
%**For other simulation results, see
%  /d/ursa/lhui/Home/Codes/Pdf/Sim/guide
%  and Dir2/ and Dir3/ for open and lambda models
%  Also, if want logarithmic binning, look at /Rs8and12/simpdfV.f code
%  simpdf.f (without the V) gives the simulation pdf with no error-bar.

%elong2.plot
%\begin{figure}[htb]
%\centerline{\psfig{figure=elong.ps,height=3.0in}}
%\caption{A ``movie'' of redshift distortions.
%At the top two panels are the positions of two representative mass
%elements
%in both real (x) and redshift-space (s), at some early
%time $t_0$. The line-of-sight is illustrated in the horizontal
%%direction. The velocity vectors
%(v (1+z)/H) are shown for the map in real-space (see eq.
%[\protect{\ref{s}}]). A density enhancement 
%between the two elements is pulling them together. Notice
%how the two points appear closer in redshift-space than in real-space.
%This is the Kaiser compression effect. 
%At some latter time $t_1$, however, 
%even though real-space multiple-streaming has not yet occurred,
%the two elements have already streamed past each other in
%redshift-space. In fact, at this particular moment,
%the distances between them in x-space and s-space are the same i.e.
%the Kaiser compression is approximately canceled. Shown at time $t_2$ is
%yet another example of translinear distortions: redshift-space
%multiple-streaming causes
%the structure to appear more elongated in s-space.
%Note that this phenomenon is physically different from the
%finger-of-God, which occurs after multiple-streaming
%occurs in real-space, and is due to virialized motions induced after
%the collapse of structures in real-space. The latter is illustrated in
%the bottom panels.}  
%\label{elongation}
%\end{figure}

\end{document}